\documentclass[a4paper, authoryear, 12pt]{elsarticle}

\usepackage{amssymb,amsmath,amsfonts}
\usepackage[ansinew]{inputenc}
\usepackage{verbatim, graphicx, ulem}
\usepackage[english]{babel}
\usepackage{amsmath, amsthm, amssymb}
\usepackage{color}
\usepackage{pgf}
\usepackage{tikz}

\usetikzlibrary{arrows,automata}

\newcommand{\beq}{\begin{equation}}
\newcommand{\eeq}{\end{equation}}
\newcommand{\bdm}{\begin{displaymath}}
\newcommand{\edm}{\end{displaymath}}
\newcommand{\bea}{\begin{eqnarray}}
\newcommand{\eea}{\end{eqnarray}}

\newcommand{\benum}{\begin{enumerate}}
\newcommand{\eenum}{\end{enumerate}}
\newcommand{\bit}{\begin{itemize}}
\newcommand{\eit}{\end{itemize}}
\newcommand{\bdes}{\begin{description}}
\newcommand{\edes}{\end{description}}
\newcommand{\bpic}{\begin{picture}}
\newcommand{\epic}{\end{picture}}
\newcommand{\bc}{\begin{center}}
\newcommand{\ec}{\end{center}}

\newcommand{\bq}{\begin{quote}}
\newcommand{\eq}{\end{quote}}

\begin{document}

\begin{frontmatter}

\title{The Significance of Non-Empirical Confirmation in Fundamental Physics}

\author{Richard Dawid}

\address{Stockholm University\\
Department of Philosophy\\
10691 Stockholm, Sweden\\
email: richard.dawid@philosophy.su.se}

\date{\today}



\begin{abstract} 
In the absence of empirical confirmation, scientists may judge a theory's chances of being viable based on a wide range of arguments. The paper argues that such arguments can differ substantially with regard to their structural similarly to empirical confirmation. Arguments that resemble empirical confirmation in a number of crucial respects provide a better basis for reliable judgement and can, in a Bayesian sense, amount to significant \textit{non-empirical} confirmation. It is shown that three kinds of non-empirical confirmation that have been specified in earlier work do satisfy those conditions. 
\end{abstract}

\end{frontmatter}

\newpage

\section{Introduction}

Fundamental physics today faces the problem that empirical testing of its core hypotheses is very difficult to achieve and even more difficult to be made conclusive. 
This general situation is likely to persist for a long time. If so, it may substantially change the perspective on theory assessment in the field. 

During most of the 20th century, fundamental physics was perceived as a scientific field where theories typically could be empirically tested within a reasonable time frame.  In high energy physics, collider experiments provided a steady and focussed stream of empirical data that could be deployed for testing physical hypotheses. In cosmology, the empirically confirmed principles of general relativity were clearly distinguished from the speculations of cosmological model building that generated a multitude of competing ideas and models without seriously insisting on the correctness of any of them at the given point. 

In this overall scientific climate, it was plausible to focus on empirical confirmation as the only reliable basis for assessing a theory's viability.  It seemed to make little sense to enter a detailed analysis of the degree of trust one was allowed to have in a theory on the basis of non-empirical evidence if conclusive empirical evidence that would decide the fate of the theory could normally  be expected to be just a few years ahead. And even if some specific cases might have rendered such an analysis interesting, the general character and status of fundamental physics seemed well understood without it.

Today, the situation is very different. String theory has been playing the role of a well established approach towards a universal theory of all interactions for over three decades and is trusted to a high degree by many of its exponents in the absence of either empirical confirmation or even a full understanding of what the theory amounts to. Cosmic inflation is being trusted by many theoreticians to a degree that in the eyes of many others goes substantially beyond what is merited by the supporting empirical data. Multiverse scenarios in the eyes of critics raise the question to what degree they can be endorsed as scientific hypotheses at all, given that their core empirical implications to a large extent seem not empirically testable in principle. What is at stake here is the understanding physicists have of the status of the theory they work on throughout their lifetimes. In the most far-reaching cases it is the status a given theory can acquire at all.    

The question as to how much credit can or should be given to non-empirical theory assessment therefore has turned from a fringe topic in physics into a question at the core of the field's self-definition. In each individual case that question has to be answered based on a detailed physical analysis of the merits and problems of the theory under scrutiny. There is a sense in which the general issue of what can be accepted as a solid basis for theory assessment is nothing more than the sum of the individual physical assessments of theories and their specific claims. 

However, the physicists' views of the scientific contexts they encounter sediment into a generalized background understanding of how theory assessment should work in a scientific way. This "philosophical background" then exerts an implicit influence on the individual physicist's theory evaluations. In periods when significant changes of the overall research context induce shifts with respect to the described sedimented philosophical background, the scientist's instrumentarium of theory assessment becomes less stable and more controversial. There are clear indications that fundamental physics today finds itself in a situation of that kind. 

By elucidating general characteristics of the ongoing shifts and making them explicit, philosophical analysis can, I believe, contribute to the process of developing an altered general understanding of theory assessment that is adequate to fundamental physics under the new circumstances. 

In Dawid (2013), it was argued that a considerable degree of trust in an empirically unconfirmed theory could be generated based on 'non-empirical theory confirmation'. Non-empirical confirmation denotes confirmation by evidence that is not of the kind that can be predicted by the theory in question, i.e. that does not lie within the theory's intended domain. 

Let me, at this point, just give a first rough idea of the difference between empirical and non-empirical confirmation. If string theory ended up making specific quantitative predictions, data in agreement with those predictions would lie within the theory's intended domain and therefore amount to empirical confirmation. In the absence of empirical confirmation, exponents of the theory may rely on different kinds of reasoning. For example, they may argue that the theory is supported by the striking difficulties to come up with promising alternatives. Those difficulties clearly cannot be predicted by string theory itself. The observation that those difficulties exist is a contingent observation about the research process of which the development of string theory is just one part. Therefore, this observation does not constitute evidence within string theory's intended domain. If one concludes, as I will, that the observation amounts to confirmation of string theory nevertheless, it can only be non-empirical confirmation. Much of the paper will be devoted to making the concept of non-empirical confirmation more precise than this rough sketch.   

Dawid (2013) spells out three specific 'non-empirical' lines of reasoning that play an important role in  generating trust in string theory. It is argued that those strategies, while playing a particularly strong role in today's fundamental physics for a number of reasons, have always constituted an important element of physical reasoning whose significance had been neglected or underrated by philosophers of science as well as by many scientists. 

The present article will look at the issue of non-empirical confirmation from a slightly different angle. Rather than discussing the conceptual details of individual arguments, it will, step by step, develop the conceptual framework for taking non-empirical confirmation seriously  at all. Sections 2 to 4 will spell out what we can plausibly hope for when assessing non-empirical arguments in favour of a scientific theory. Section 5 will then develop guidelines for identifying arguments of non-empirical confirmation that can fulfil those expectations. Section 6, finally, will demonstrate that the three arguments of non-empirical confirmation presented in Dawid (2013) are promising for the very reason that they do satisfy the conditions developed in Section 5. This fact, it shall be argued, offers a plausible reason why the three presented arguments may be taken to be more powerful than other reasons for having trust in a theory that might come to mind. The three arguments themselves will only be rehearsed briefly in Section 6. The reader interested in a more careful presentation and discussion of those arguments may look into Dawid (2013), Dawid, Hartmann and Sprenger (2015) and Dawid (2016).


\section{Strategic or Epistemic?}

It is uncontroversial that physicists assess theories already prior to empirical testing. But what kind of question do they ask when making those assessments?
In the previous section, it has been claimed that many exponents of string theory, cosmic inflation and multiverse scenarios have generated a degree of trust in their theories that cannot be explained entirely - or, in the case of string theory, cannot be explained at all - by empirical confirmation of those theories. It was thereby implicitly taken for granted that scientists aim at assessing the degree of trust they should have in a theory's truth or viability.

One might take one step back, however, and consider a more restrained, and therefore possibly less contestable point of view. On that view, the crucial question for the working scientist is simply whether or not it makes sense to work on a given theory. Endorsing a theory in this light might be taken to be the result of fairly pragmatic considerations that do not address the more ambitious question whether or not a theory is likely to be true or viable.

While I concur that questions of research strategy are a main motivating force behind theory assessment, I do not think that what is at stake in non-empirical theory assessment can be reduced to the pragmatic issue of deciding upon research strategies. 

Clearly, a scientist can have a number of rational reasons for working on a given scientific theory that are entirely unrelated to the question of the theory's viability. To begin with, there can be personal strategic reasons. A young physicist might decide to work on the most popular theory because this seems most promising with respect to job perspectives; or she might decide to work on a nascent less developed theory because that seems most promising with respect to assuming a leading role in a research field. Other reasons address what is in the best interest of the scientific discipline. It may be most effective to first work out a theory that has already been proposed before starting a tedious search for possible alternatives. Among a spectrum of known approaches to solving a problem, it may be reasonable to first work out the approach that is easiest to develop. Even though none of these considerations addresses the question whether or not a given theory is likely to be viable, they can all make sense and do play a role in physics.

However, their relevance for deciding on the optimal research strategy in a scientific field is limited for two reasons. First, one often finds lines of reasoning that pull in different directions and whose balance changes in time. For example, the idea that it is most effective to work on a theory that has already been developed may be countered by the idea that it is sometimes more productive to start from scratch rather than to be bound by worn-out ways of thinking. Second, and maybe even more importantly, entirely strategic considerations on theory preference look plausible only if based on the assumption that all alternative hypotheses are about equally promising. Any deviation from that assumption draws into question the implications of considerations that disregard epistemic issues.  

The strongest reasons for working on a theory in this light are those that do have an epistemic foundation suggesting that a theory is likely to be viable. To the extent epistemic arguments can be developed, it is of high importance for the physicist to take them into account.  
The physicist herself may well treat those arguments pragmatically as a way of understanding whether there are good reasons to work on a given theory. From an operative professional perspective, framing what is at issue in that way is perfectly adequate. But in the end, a strong commitment to working on a particular theory hinges on the question whether there exists a good reason for having trust in the theory's viability. 

I want to point out a second reason for emphasizing the epistemic element in non-empirical theory assessment.  Reducing what is at stake in non-empirical theory assessment to the question of justifying work on a theory seems at variance with the main motivation for doing fundamental physics. Fundamental physics today clearly is not driven by perspectives of technological utilization. It is driven by the quest for acquiring knowledge about the world. In fields aiming for technological utilization, it is plausible that non-empirical theory assessment is primarily motivated by the strategic question as to whether one should focus on a given research program. When medical researchers look for a cure for a disease, assessments of a specific approach may be motivated entirely by the need to decide whether one should work on that approach or on a different one. All that counts is the final result of having a cure that is sufficiently well tested to use. Assessing the truth of a hypothesis is of little relevance as long as it does not open up a perspective of utilization in the foreseeable future.

It is important to distinguish that situation from the situation fundamental physics finds itself in today. In a research field that is motivated primarily by the quest for knowledge, understanding the probability of a theory's viability is not just a pragmatic step on the way towards conclusive confirmation, it is also an epistemic goal in its own right. Knowing that there is a high probability that dinosaurs got extinct due to a comet impact constitutes a valuable element of my knowledge about the world even though that hypothesis has not been conclusively confirmed. It would remain valuable even if there was no hope of conclusive confirmation in the foreseeable future. Similarly, to the extent there are reasons to assume that string theory or the multiverse have a high chance of being viable, that constitutes an important element of knowledge about the universe even in the absence of conclusive empirical confirmation of those theories. 

It may make sense to ignore "intermediate" epistemic states between ignorance and conclusive knowledge in contexts where they last only for a brief period of time before the case is settled based on conclusive empirical evidence. In contemporary fundamental physics the typical time scale for that intermediate state has grown beyond one generation of scientists. In  such a situation, ignoring the epistemic difference between knowing nothing about a unified description of nuclear interactions and gravity and, to the extent that can be established, knowing that string theory is likely to represent such a description amounts to a substantial misrepresentation of present day knowledge about fundamental physics. 

\section{Confirmation}

\subsection{Bayesian confirmation is not conclusive confirmation}

In light of the previous arguments, the crucial question is: can there be strong reasons for believing that a theory is probably viable even if that theory has not found empirical confirmation? The natural framework for discussing this question in the philosophy of of science is provided by Bayesian confirmation theory. Bayesian confirmation theory expresses confirmation in terms of the theory's probability, which is exactly the kind of perspective that allows for a specific analysis of the question stated above. According to Bayesian confirmation theory, evidence $E$ confirms a hypothesis $H$ iff the posterior probability of $H$ given evidence $E$ is higher than the prior probability of $H$:

\beq
P(H|E) > P(H)
\eeq  

In Bayesian wording, we therefore want to understand whether and to what degree non-empirical theory assessment amounts to theory confirmation: we are looking for \textit{non-empirical theory confirmation}. 

It has been criticized (see e.g. Ellis and Silk 2014, Rovelli 2017) that the term \textit{non-empirical confirmation} suggests that the given theory has been established as viable and is in no need of further empirical testing. It is important to emphasize that the Bayesian definition of confirmation does not imply anything of that kind.
It is true however, that  the communication between theoretical physics and the philosophy of science on this issue is made difficult by an unfortunate mismatch between the ways the term "confirmation" is used in the two fields. 

In statistics and in parts of experimental physics that engage in statistical data analysis, the use of the term "confirmation" is confined to Bayesian data analysis, where it denotes the increase of probability described above. In contemporary philosophy of science, confirmation is mostly understood in terms of Bayesian confirmation theory as well. While some important voices in the philosophy of science (see e.g. Achinstein 2001) disagree with a Bayesian understanding of confirmation, the Bayesian approach is dominant in the field to the extent that the term "confirmation" in the title of a paper in the philosophy of science without further qualification amounts to the announcement that the paper is written within the framework of Bayesian epistemology. For that reason, a philosophical paper on the issue needs to refer to a Bayesian definition of confirmation in order to specify for a philosophical audience what is being discussed. 

In theoretical physics, however, the term confirmation is often used in accordance with its non-technical use in the sense of what one might call "conclusive confirmation": the theory has been established to be viable in a given regime beyond reasonable doubt. Due to those two conflicting uses of the term "confirmation", the term "non-empirical confirmation" can indeed cause misunderstandings for listeners who are not familiar with the Bayesian use of the term. 

Let me thus state unequivocally: non-empirical confirmation does not mean conclusive confirmation.  Readers who for terminological reasons feel uncomfortable with the term "non-empirical confirmation" might, before their mental eye, replace it with the expression "non-empirical theory assessment" with the asterisk that "assessment" is to be understood in terms of attributing a probability of being viable to the theory in question.  

\subsection{Viability rather than truth}

This leads to a second important specification of the way confirmation is used in the present context. The previous paragraph defined confirmation in terms of a theory's viability. The canonical formulation of Bayesian confirmation, however, is based on a theory's truth probability. The fact that I deviate from this canonical approach is closely related to the role confirmation is going to play in this analysis and deserves closer attention. 

Truth is a complicated concept. Whether or not a mature and empirically well confirmed theory may be taken to be true or approximately true has been hotly debated in the philosophy of science for a long time. Things are particularly difficult in theoretical physics where some theories are predictively highly successful even though they are known not to be true for conceptual reasons. Quantum field theories are predictively highly successful based on the first orders in a perturbative expansion that is strongly suspected not to converge and therefore difficult to appraise in terms of truth values. The standard model of particle physics is highly successful but known to constitute merely an effective theory to whatever more fundamental theory can account for the inclusion of gravity at the Planck scale. 

Many Bayesian epistemologists are not troubled by the notorious difficulties related to calling scientific theories true for one reason: Bayesian confirmation theory relies on a differential concept of confirmation. Evidence confirms a hypothesis if it increases its truth probability. The verdict that it does so in a given scenario is invariant under the choice of priors for the hypothesis as long as one excludes the dogmatic priors zero and one. All doubts about the the truth of scientific theories can be relegated to specifying the priors, however. Therefore, even if one attributes a truth probability very close to zero to a scientific theory for philosophical reasons, one can still talk about the formal increase of a theory's truth probability based on new evidence.

While this line of reasoning works sufficiently well in contexts of empirical testing, it seems less than satisfactory in the context of investigating the role of non-empirical confirmation in fundamental physics. First, as noted above, the understanding that some predictively very successful theories in fundamental physics are strictly speaking not true is related to the understanding that they are, strictly speaking, inconsistent. If so, it seems adequate to attribute probability zero to their truth, which would bar any meaningful updating on truth probabilities.     

Second, and more directly related to the agenda of this paper, defining Bayesian confirmation in terms of truth probabilities has an important effect on the ways confirmation can be discussed. The analysis above already stated the main reason why, according to the understanding of most Bayesian epistemologists, not even a rough consensus on absolute values for truth probabilities can be established empirically: different philosophical views may suggest very different truth probabilities even once a theory is empirically very well confirmed. 

This implies, however, that a Bayesian analysis of confirmation in terms of truth probabilities will only be convincing to the extent it avoids any reference to absolute probabilities. Doing so can work to a given extent with respect to empirical confirmation if one is willing to decouple the issue of the significance or conclusiveness of evidence from the issue of absolute probabilities of the theory. Whether or not a theory can be called conclusively tested is then taken to be decided by the involved scientists based on specifying significance criteria for p-values within the framework of frequentist data analysis.\footnote{Strictly speaking, this is philosophically satisfactory only once one has spelled out the connection between p-values and truth probabilities. I won't address that issue here, however.}

Non-empirical theory confirmation, however, is not based on a solid setup of well specified rounds of testing that can aim at surpassing strict significance limits. The issue of the significance or conclusiveness of non-empirical evidence therefore must be decided by assessing absolute values of probabilities.\footnote{As always in Bayesian epistemology, this does not mean that we aim at extracting actual numbers. But the aim must be to demonstrate that fairly high absolute probabilities can be made plausible in the process.} If so, however, it is important to specify what is denoted by the probabilities in a way that is philosophically uncontroversial and can be linked to empirical data. Truth, for the reasons spelled out above, fails to meet those conditions. 

I therefore propose to understand confirmation as an observation-based increase of the probability that the theory is viable in a given regime. Viability here is defined as the agreement of the theory's predictions with all empirical data that can be possibly collected within a given regime. Regimes of empirical testing are specified based on well established background knowledge about a given research context. An example of specifying a regime of testing in the context of high energy physics would be specifying a certain energy scale up to which a theory is tested. The theory is viable within that regime if it correctly accounts for all possible empirical testing up to the specified energy scale.      

The probability of a theory's viability therefore can be specified only with respect to a certain regime of testing.

\section{Significant non-empirical confirmation?}

\subsection{There is no plausible claim of conclusive non-empirical confirmation}

How strong should claims of non-empirical confirmation, understood in the sense spelled out above, be taken to be? It has been emphasized in Section 3.1. that the Bayesian concept of confirmation does not \textit{imply} conclusive confirmation. But could non-empirical evidence get so strong that we had to accept it as conclusive? 

The conceptual status of non-empirical confirmation is somewhat reminiscent of the conceptual status of third person testimony. There is an irreducible difference between observing something with one's own eyes and learning about it from a third person. The best way to check a given testimonial is to have a look with one's own eyes. Still, a sufficiently dense web of reliable testimony for a given fact can provide a basis for taking that fact for granted. (Science crucially relies on the fact that this is possible.) Whether or not a sufficiently reliable web of testimony can be attained is a contingent fact about the world or, more specifically, a given context of inquiry. 

In a similar way, there exists an irreducible difference between empirical and non-empirical confirmation. The best way to check whether non-empirical confirmation works is to control it based on empirical testing. Whether or not one is willing to accept non-empirical evidence as conclusive depends on the reliability of this kind of confirmation in the past. Let us assume, for a moment, that we lived in a world where a given strategy of non-empirical confirmation was omnipresent and had a 100\% success rate over many generations. In such a world, it would be just as irrational not to rely on non-empirical evidence of that kind as it is in ours to refuse stepping on bridges because of the Humean problem of induction. Understanding the strength as well as the limitations of non-empirical confirmation is itself a matter of observing the world.

The strength  non-empirical confirmation could acquire in principle needs to be clearly distinguished from the strength one attributes to non-empirical confirmation in actual physics. The actual power of non-empirical confirmation in science is limited. No case of conclusive non-empirical confirmation has ever occurred in any field of science up to this day. A number of factors that will emerge later in the discussion reduce the reliability of non-empirical confirmation in a way that keeps it far from being conclusive even under optimal circumstances. 
Given our present understanding of the scientific process, it cannot be expected that non-empirical confirmation will lead to the conclusive confirmation of a hypothesis in physics in the future. Strong empirical testing therefore must be expected to remain the only path towards conclusive confirmation of a theory in fundamental physics.

\subsection{But one needs to establish significant non-empirical confirmation}

So nothing in this paper will aim at suggesting that non-empirical confirmation be conclusive. Still, it will be important to claim something stronger than the mere point that non-empirical evidence amounts to confirmation in a Bayesian sense at all.  Though by no means trivial, the latter claim would be too weak to be interesting.
For the Bayesian, even a minimal increase in probability amounts to confirmation. By establishing that a given line of reasoning amounts to non-empirical confirmation, nothing has been said about the significance of that kind of confirmation. Therefore, the mere fact that non-empirical evidence can confirm does not establish that it can play a relevant role in scientific reasoning. In order to support the latter claim, one needs to make plausible that non-empirical confirmation is significant in the sense that it can lead to substantial probabilities for a theory's viability even when starting with low initial expectations.\footnote{In the context of a Bayesian formalization, posterior probabilities are always a function of the subjective priors. In any specific case, a high posterior probability thus hinges on a prior that is not too low. In the formalized model, the statement that some evidence provides significant confirmation therefore does not guarantee a high posterior. It merely implies that high posteriors are reached based on the given evidence from a wide range of priors.} It has already been pointed out that the need to look at absolute probabilities in order to establish significant confirmation is one crucial reason for defining confirmation in terms of a theory's viability rather than its truth. 

The distinction between plain confirmation and significant confirmation will play an important role in our analysis once the specific strategies of non-empirical confirmation have been spelled out. It will be argued that each of the individual strategies of non-empirical confirmation in isolation does constitute confirmation but cannot be established to be significant. As we will see, significant confirmation can only be made plausible based on two or even three arguments of non-empirical confirmation in conjunction.

\section{Zooming in on Non-Empirical Confirmation}

Let us recapitulate what we are looking for. We want to identify strategies of attaining significant non-empirical theory confirmation. That is, we are looking for kinds of observation that lie beyond a theory's intended domain (which means that they are not of the kind that can be predicted by the theory) but can nevertheless substantially increase the theory's probability of being viable. 

A wide range of arguments may be taken to increase a physicist's trust in a theory. (a) One may point to a theory's elegance, beauty or simplicity and argue that a theory that shows those qualities to such a high degree must be expected to be viable. (b) One might point to one's gut feeling (or to the gut feeling of some very prominent physicist) and argue that a very strong gut feeling about the theory's viability indicates that the theory is probably viable. (c) A recent suggestion has been that one might trust a theory based its mathematical fertility (Rickles 2013.)
Arguments of the described kinds clearly do influence scientists' expectations regarding a theory's viability. There is no reason to dogmatically deny that such considerations can indeed be helpful in certain contexts.

However, none of the lines of reasoning just listed provide a convincing basis for reconstructing them as inter-subjectively viable and therefore genuinely scientific strategies of theory assessment. I will briefly address suggestion (c) a little later. At this point, I want to focus on suggestions (a) and (b), which share two substantial flaws. 

First, there are no clear parameters for measuring a theory's elegance, beauty or simplicity, let alone for measuring the intensity of a scientist's gut feeling. Any attempt to infer a theory's viability from those qualities thus looks hopelessly vague and subjective. 

Second, even if there were a way of making the attribution of elegance or similar qualities more precise, the way the argument is set up confines it to evaluating characteristics of the theory itself and drawing inferences from that evaluation. At no point does the argument reach out to an observation about the actual world. This limited scope of the argument creates a serious problem, however. Inferring a theory's viability from internal features like elegance, beauty, etc.  must rely on the understanding that the given features are conducive to that theory's viability. If no observations about the external world are involved in the argument, the claim about the connection between the features in question and the theory's viability cannot be treated as empirically testable but must be accepted as a dogmatic posit. An inference from those features to the theory's viability on this basis cannot be accepted as scientific reasoning. 

In a nutshell, one might characterize the status of the discussed arguments the following way. They are epistemic in nature because they do address a theory's prospects of being viable. The vagueness and subjectivity of the criteria applied and the lack of explicit connections to observations about the external world implies, however, that the arguments' significance cannot be established based on a concise and scientifically legitimate line of reasoning.
An individual scientist may act based on the suspicion that the way she understands and deploys the arguments guides her towards theories with better prospects of success. This may be sufficient for justifying the use of those arguments when deciding which way to go in the absence of stronger arguments. It does not allow for an intersubjective justification of substantial trust in a given theory, however. The described arguments don't amount to substantial scientific theory confirmation.

Adherents to a canonical understanding of theory confirmation would assert that this is how far one can get in the absence of empirical confirmation. By presenting the case for non-empirical theory confirmation, I contest that view. I assert that specific arguments of non-empirical confirmation are substantially stronger than the lines of reasoning analysed so far. They are stronger because their argumentative structure lies closer to the rationale of empirical confirmation in important respects. The next question we need to address is: which conditions should be fulfilled by a promising candidate for substantial non-empirical confirmation?  
In the following, I will present and motivate three conditions. 

\subsection{Observations about the world}

The first criterion for promising non-empirical theory confirmation can be extracted directly from the previous discussion.  We need to distinguish between the system made up of the scientists and the theories they have developed, on the one hand, and the world beyond that system on the other. In order to amount to a scientifically legitimate kind of reasoning, non-empirical confirmation must be based on observations about the \textit{external} world beyond the system made up of scientists and their theories. 

With respect to a given theory, observations about that \textit{external} world can in turn be divided into two parts. First, there is the intended domain of the theory, consisting of observations of the kind that can be predicted by the given theory. And second, there are observations beyond that intended domain. Observations in the intended domain can be confronted with the theory's predictions and on that basis can provide empirical confirmation or dis-confirmation. Observations about the external world that lie outside a given theory's intended domain cannot contribute to empirical testing of that theory but may nevertheless be relevant for assessing the theory's viability. They are candidates for non-empirical confirmation.

\subsection{A predictive meta-level hypothesis}

The second condition to be met by a convincing argument of non-empirical confirmation is related to the mechanism of confirmation itself. In order to speak of confirmation, we need a clear argument why the confirming evidence increases the probability of the viability of $H$.
In the case of empirical confirmation, this probability increase can be immediately understood. For the sake of simplicity, I spell it out for the case of a deterministic theory where data $E$ is strictly implied by theory $H$. In other words, we have $P(E|T)=1$. The core argument remains unchanged in the case of probabilistic predictions.
Using $P(E|T)=1$ and the total probability $P(E)=P(T)P(E|T)+P(\neg T)P(E|\neg T)$,  we can write the Bayes formula as

\bea
P(T|E) & = & \frac{P(E|T)P(T)}{P(T)P(E|T)+P(\neg T)P(E|\neg T)}  \\
 & = & \frac{P(T)}{P(T)+P(\neg T)P(E|\neg T)}
\eea

If we exclude $P(T)=1$ and $P(E|\neg T)=1$, which are the trivial cases that we are sure either about the viability of $H$ or about the outcome of the experiment already before measurement, we get $P(T|E)>P(T)$, which means that $E$ confirms $H$. In effect, $H$ gets confirmed by $E$ because measuring $E$ sets to zero the probabilities of potential (conceived or unconceived) alternative theories that are in conflict with $E$. 

Non-empirical confirmation is based on data $F$ that lies beyond the theory's intended domain. Therefore, the confirmation mechanism that applied in the case of empirical confirmation is absent. How can we replace it? How can we make plausible that non-empirical evidence $F$ increases the probability of the viability of theory $H$?
The most straightforward strategy is the following: try to find a construction that structurally resembles the one that applies in empirical confirmation. That is, isolate a hypothesis $Y$ that is predicted by data $F$. This means that $F$ confirms $Y$. The crucial question now becomes whether the viability or truth of $Y$ is positively correlated with the viability of $H$. If that is the case, the increase of the probability of $Y$ that is induced by $F$ feeds down to $H$. $F$ then amounts to non-empirical confirmation of $H$. 

$Y$ must not be expected to be a genuine scientific theory that can be empirically tested in a scientific sense. The way in which $F$ confirms $Y$ will be more vague, which is one core reason why non-empirical confirmation should not be expected to be conclusive. However, it shall be argued, a vague relation between $Y$ and $F$ can be sufficient for generating quite substantial trust in a scientific theory $H$ if $F$ is sufficiently well specified.

What we are thus looking for is a meta-level mechanism that resembles empirical confirmation and feeds down to the ground level  hypothesis. An argument of that form will arguably be the closest one can get to empirical confirmation in the absence of empirical data in the theory's intended domain.

It should be emphasized that presenting a convincing meta-level hypothesis $Y$ is by no means an easy task. In order to see this, let us for a moment return to the claim, put forward in Rickles (2013), that a theory's "mathematical fertility", i.e. its tendency to lead to the development of new mathematics, increases that theory's probability of being viable. Observing the mathematical fertility of a physical theory may be understood in terms of a contingent observation about the research process and therefore satisfies condition 5.1. Based on the considerations presented in this subsection, it could therefore provide the basis for a promising argument of non-empirical confirmation if a meta-level hypothesis $Y$ could be found that predicted mathematical fertility and was positively correlated with the theory's viability. 

It seems quite tricky, however, to come up with any hypothesis of this kind. First, it seems complicated to formulate any hypothesis at all that predicts the mathematical fertility of a scientific theory. And second, in order to be conducive to a theory's viability to a significant degree, that hypothesis would need to screen off those contexts of reasoning that are mathematically fertile by actually \textit{reducing} the chances of a theory's viability. For example, if the world's complexity was understood to lead to mathematical fertility of theories that can describe it, it might seem even more fertile to think about a world that is more complex than ours. Thinking about a world of very many spatial dimensions arguably was mathematically fertile for a 19th century mathematician for the very reason that it moved beyond what seemed physically realistic. So the understanding that the world's complexity predicts mathematical fertility of theories does not generate non-empirical confirmation value of the observation of mathematical fertility because describing more complexity than the world contains may seem mathematically even more fertile. In order to establish a significant confirmation value of mathematical fertility, one would have to find meta-level hypothesis $Y$ that was clearly positively correlated with a theory's viability and predicted mathematical fertility.  I don't want to exclude that this is possible. But it is by no means a foregone conclusion that a plausible $Y$ can be found in the given case. In the absence of a known hypothesis $Y$ that fits the bill, the given argument's significance remains doubtful.

\subsection{Viability rather than truth (again)}

A third consideration is important to emphasise. In recent years, it has been a matter of debate in some parts of fundamental physics whether a physical theory could be called scientific if it did not make any empirical predictions even in its fully developed form. While thinking about non-empirical confirmation might open up new perspectives on this debate, one would not want to set up non-empirical confirmation in a way that implies a positive answer to this highly contentious question. As pointed out earlier in this section, it is crucial for establishing  the plausibility of non-empirical confirmation to exploit all available connections to empirical data. The more non-empirical confirmation gets decoupled from observation, the more questionable its implications will become. In this spirit, the goal of making the strongest possible case for non-empirical confirmation suggests to confine the analysis to theories that  do have empirical implications. We do want to account for theories like string theory that are not sufficiently understood at this point for extracting specific empirical predictions. It shall be assumed, however, that theories under consideration are expected to reveal specific empirical implications once sufficiently understood.

Confining our analysis to empirically relevant theories leads us back to the choice between addressing a theory's truth and its viability that has already been discussed in Section 3.2. It offers one more strong reason for preferring the concept of viability over the concept of truth. If we understood confirmation in terms of an increase of truth probability, it would be difficult to exclude cases of theories that are predictively empty. A theory's truth may be taken to be distinct from its empirical adequacy, which means that a non-empirical argument could in principle increase a theory's truth probability without changing that theory's probability of being empirically adequate. This opens up the possibility of confirming predictively empty theories.

Empirical viability, to the contrary, has been defined in terms of the theory's empirical predictions. A theory is viable in a given regime if its predictions agree with all data that can in principle be collected in that regime. If the theory is predictively empty in that regime, its viability in that regime is trivial and non-empirical evidence plays no role in establishing it. Non-empirical confirmation defined in terms of viability therefore is applicable only to theories that are predictive in principle.

\vspace{5mm}

We have now specified three conditions that should be fulfilled by non-empirical confirmation of a theory $H$ in order to make it as similar to empirical confirmation as possible. 

\vspace{5mm}

{\bf 1)} The observations on which non-empirical confirmation is based should be about the external world rather than merely about the system of scientists and their theories.

\vspace{5mm}

{\bf 2)} It should be possible to construe an argument of non-empirical confirmation  based on "soft" empirical confirmation of a meta-level hypothesis $Y$ by well-specified non-empirical evidence $F$ which, in conjunction with a positive correlation between $Y$ and the viability of $H$, establishes that $F$ confirms $H$.

\vspace{5mm}

{\bf 3)} Non-empirical confirmation should be applicable only to empirically predictive theories, which is guaranteed by defining confirmation in terms of the theory's viability.

\vspace{5mm}

An argument of non-empirical confirmation that meets  those three conditions has a plausible path towards being significant.

\section{A specific realization}

We are now in a position to connect to the three core strategies of non-empirical confirmation that have been presented in Dawid (2013). As will be shown in this section, they all do satisfy the three conditions specied in the previous section.
Let me restate the three arguments of non-empirical confirmation.

\vspace{5mm}

{\bf NAA:} The No Alternatives Argument:  Scientists have looked intensely and for a considerable time for alternatives to a known theory $H$ that can solve a given scientific problem but haven't found any. This observation is taken as an indication of the viability of theory $H$.

\vspace{5mm}

{\bf MIA:} The Meta-Inductive Argument from success in the research field:
Theories in the research field that satisfy a given set of conditions have shown a tendency of being viable in the past. This observation is taken to increase the probability that a new theory $H$ that also satisfies those conditions is also viable.
\vspace{5mm}

{\bf UEA:} The Argument of Unexpected Explanatory Interconnections:
Theory $H$ was developed in order to solve a specific problem. Once $H$ was developed, physicists found that $H$ also provides explanations with respect to a range of problems which to solve was not the initial aim of developing the theory. 
This observation is taken as an indication of the theory's viability.

\vspace{5mm}
 
Since the arguments are formulated as inferences to a higher probability of the theory's viability, they by definition satisfy condition three. It is also straightforward to see that condition one is satisfied, since for all three arguments the respective observation about the external world has been spelled out in the first sentence. This leaves us with the task to understand that the arguments are of the form required by condition two. Since Dawid (2013) contains an extensive analysis of this point, it will suffice in the present paper to give a brief sketch of the line of reasoning. The reader interested in a more through discussion is referred to Dawid (2013), Chapters 2.2  - 3.2.

The meta-level hypothesis $Y$ that seems most adequate in all three presented arguments of non-empirical confirmation is a statement on local limitations to scientific underdetermination. 
In order to understand that statement we need to clarify the terminology. 

Scientific underdetermination (called transient underdetermination in a slightly different context by Lawrence Sklar (1975) and Kyle Stanford (2006)), measures how many alternative theories that are not empirically fully equivalent to each other (that is, they could be distinguished empirically in the long run) can account for a given empirical data set. 

Local scientific underdetermination only accounts for theories that could be distinguished within a given "experimental horizon", that is, a specified class of experiments. For example, local scientific underdetermination up to a given energy scale in high energy physics accounts for the spectrum of possible theories that may have different empirical implications up to that energy scale. The concept of local scientific underdetermination therefore stands in direct relation to a theory's viability within a given experimental horizon as introduced in Section 3: local scientific underdetermination with respect to a given empirical horizon counts those theories that can have different empirical viability conditions with respect to that empirical horizon.    

Limitations to scientific underdetermination denote constraints on scientific underdetermination of some kind. Such limitations are a natural candidate for $Y$ for a simple reason: any trust in predictions of an \textit{empirically confirmed} scientific theory must be based on implicit assumptions of limitations to scientific underdetermination. 

If scientific underdetermination were entirely unlimited, one had to expect that any imaginable continuation of the observed regularity patterns could be accounted for by a plausible scientific theory. There would be no basis for expecting that, among the unconstrained mass of such theories, it was precisely the one scientists happened to have developed that was empirically viable. Only a rather radical rejection of the idea of unlimited underdetermination can lead physicists towards trusting the predictions provided by their theory. 
In other words, any degree of trust in a theory's viability in a fairly straightforward way translates into the meta-level hypothesis that local scientific underdetermination is strongly limited. 

A very high degree of confidence in a theory actually translates into the understanding that, within the considered empirical horizon, there probably is no alternative scientific theory at all that can account for the available data in a satisfactory way but makes different predictions within the empirical horizon. If there were such unconceived alternatives, how should one rule out that one of those rather than the known theory was the viable one?

The question remains: how can scientists infer hypotheses on the spectrum of possible alternatives at all?
Obviously, such meta-level hypotheses cannot be supported by the empirical data that supports the scientific theory under scrutiny: that data would support any unconceived alternative as well. Support of the meta-level hypothesis therefore needs to come from observations beyond the theory's intended domain: observations of the very kind that amount to non-empirical confirmation. We remember from Section 5.2 that a confirmatory value of $F$ for $Y$ can be established by showing that $Y$ predicts $F$ (makes $F$ more probable).

Looking at the three kinds of observations $F$ listed above, we find that they all indeed do confirm a hypothesis $Y$ on limitations to local scientific underdetermination on those grounds.

Most straightforwardly, the NAA observation that scientists have not found alternative theories is predicted by a strong hypothesis Y. In the most extreme case, if no alternatives to theory $H$ exist, scientists cannot find any. (For a general Bayesian formalization of NAA and a proof that it amounts to confirmation, see Dawid, Hartmann and Sprenger (2015)).

The MIA observation that theories comparable to $H$ tended to be predictively successful once tested also is predicted by a $Y$ hypothesis on the given ensemble of theories: if in the cases considered there usually were no or very few possible alternatives to the theories under scrutiny, one would expect a fairly high percentage of those theories to be viable.

 The UEA case is a little less straightforward but eventually allows for the same conclusions. The basic idea is the following. Let us imagine a set of problems that need to be solved in a given context. Let us first assume that the number of theories that can be developed in that context is much higher than the number of problems. In this case, there is no reason to expect that a theory developed in order to solve one specific problem will be capable of solving other problems as well. Next, let us assume that the number of theories that can be developed in the given context is much smaller than the number of problems. In that case, one must expect that each possible theory will on average solve a substantial number of problems and therefore will show UEA. In the extreme case where only one theory can be developed, that theory must be expected to solve all problems. Therefore, $Y$ does predict UEA. 

We can conclude that NAA, MIA and UEA do rely on one common meta-level hypothesis $Y$ that provides the basis for establishing their character as non-empirical confirmation. The three arguments thus satisfy all three conditions set out in Section 5 for promising arguments of non-empirical confirmation. Moreover, the fact that all three arguments are related to the same meta-level hypothesis shows that they constitute a coherent set of arguments that may work well in conjunction. 

The last point becomes particularly important once one aims at understanding the arguments' actual significance.
What has been established up to this point is that NAA, MIA and UEA each amount to non-empirical confirmation. It has been pointed out in Section 4, however, that non-empirical confirmation can be taken to be scientifically relevant only if it can be shown to be significant.

The problem is that none of the three arguments in isolation can be established as significant confirmation. NAA faces the problem  that hypothesis $Y$, asserting strong limitations to local scientific underdetermination, is not the only hypothesis that can explain the observation  $F_{NAA}$ that no alternatives have been found. One might also propose hypothesis $U$ as an explanation, asserting that scientists just have not been diligent or clever enough to find the alternatives that exist. The significance of NAA therefore depends on the prior probabilities one attributes to $Y$ and $U$. An observer who attributes a very low prior to $Y$  and a substantial prior to $U$ will take $F_{NAA}$ to significantly increase the probability of $U$ and only find a fairly insignificant increase of the probability of $Y$.

MIA faces a different problem: observing a tendency of predictive success for a specified set of theory does not lend strong support to claims about the prospects of a new theory as long as there is no good reason to assume that the new theory shares the success prospects of the ones previously considered. 

UEA may seem a little more autarkic than the other two arguments but also remains questionable in isolation. Unexpected explanatory interconnections may be due to deeper underlying connections between the problems addressed by the theory that are not related to the specific theory itself. If so, they don't provide good arguments for trusting a specific theory.

The fact that the three arguments rely on the same $Y$ comes to the rescue in this situation. It provides the basis for strengthening one argument by another. This can eventually generate significant non-empirical confirmation of a theory based on two or three arguments in conjunction. 

For example, MIA offers exactly what is missing in NAA: it allows for discriminating between support for hypotheses $Y$ and $U$. While the observation of a tendency of predictive success in a field can be explained by a hypothesis $Y$ on limitations to underdetermination, it clearly cannot be explained by a hypothesis $U$ that states the limited capabilities of the involved scientists. 

Vice versa, NAA can provide what is missing in MIA in isolation: being itself related to $Y$, NAA provides plausible criteria for specifying the ensemble of theories one should consider when looking for a tendency of predictive success. When other theories that allowed for a NAA argument show a tendency of predictive success, that provides a plausible basis for expecting predictive success for a new theory where NAA also applies. 
NAA and MIA in conjunction therefore can generate significant non-empirical confirmation. 

In a case like string theory, the question whether the new theory (string theory) is in the same category as those (like the standard model of high energy physics) that are the prime examples of previous predictive success is particularly tricky because the new theory is so much more difficult and insufficiently understood. One might fear that the scientists' capability of assessing the spectrum of possible alternatives may be substantially reduced in this more difficult scientific environment, which would render the application of MIA questionable. In a situation of that kind, UEA  plays a crucial role. It can establish that the given research context is understood to a sufficient degree to extract the observed instances of unexpected explanatory interconnections. Therefore, apart from providing an argument for limitations to scientific underdetermination in its own right, UEA also strengthens the case for the applicability of MIA.
A case like string theory thus requires all three arguments of non-empirical confirmation in conjunction in order to generate significant non-empirical confirmation.


\section{Conclusion}

The emerging picture of non-empirical confirmation is the following. While there is no reasonable basis for expecting that non-empirical confirmation can become \textit{conclusive} in our world, it seems plausible that non-empirical confirmation can be \textit{significant} in certain contexts. 

The basis for significant non-empirical confirmation is an argumentative structure that is closely related to the structure of empirical confirmation in a number of respects. 

\begin{itemize}

\item Confirmation is confined to empirically relevant statements by defining it in terms of viability rather than truth.

\item Non-empirical confirmation is based on an observation $F$ about the external world beyond the system of the scientists and their theories. 

\item The connection between successful prediction and confirmation that characterizes empirical confirmation is, in the case of non-empirical confirmation, reflected by the fact that $F$ is predicted by a meta-level hypothesis $Y$. The confirmation value of $F$ then feeds down to theory $H$ based on a positive probabilistic correlation between $Y$ and $H$. 

\item Non-empirical confirmation is directly connected to empirical confirmation based on MIA. Significant non-empirical confirmation therefore only arises if there is empirical confirmation somewhere else in the research field. 

\item Arguments of non-empirical confirmation generate a web of interconnected reasoning. The individual arguments can strengthen each other in an effective way because they can all be construed in terms of the same meta-level hypothesis Y, the hypothesis of strong limitations to local scientific underdetermination. The very same hypothesis $Y$ plays a crucial role also for understanding the relevance of \textit{empirical} confirmation.

\end{itemize}

All these points in conjunction can make  non-empirical confirmation significant. This does not imply that non-empirical evidence for a theory should be dogmatically constrained to evidence that can be framed in precisely the suggested way. But the list of characteristics shared by NAA, MIA and UEA demonstrates that those arguments have fairly non-trivial conceptual merits. There are plausible reasons why those arguments in particular can become significant and play a leading role in supporting empirically unconfirmed theories.

How strong the web of non-empirical confirmation should be taken to be in a given case is, of course, a matter of detailed scientific analysis of a physical theory. There are examples in recent fundamental physics where very strong and nearly unanimous trust in a hypothesis emerged based on non-empirical confirmation. 
A particularly strong example is the trust in the viability of the Higgs mechanism before the discovery of a Higgs-Boson in 2012.\footnote{That trust was invested in the general characteristics of the Higgs mechanism. The specific details of the actual Higgs model, whether the Higgs-Boson was fundamental or had constituents, whether it was of a standard model type, supersymmetric, or other, remained open and in part have not been clarified up to this day.} String theory constitutes an example where a large section of the theory's experts has come to the conclusion that non-empirical confirmation is significant though not conclusive. Others doubt the significance of non-empirical confirmation in this case. Cosmic inflation is a case where supporting data for the theory exists but is not conclusive. Non-empirical confirmation arguably plays a substantial role in increasing trust in the theory in the eyes of many of its exponents.

\section*{Acknowledgements}

This paper has profited substantially from very helpful comments from the audience of the "Why Trust a Theory?" workshop in Munich and in particular from discussions with Peter Achinstein, Radin Dardashti, Daniele Oriti, Carlo Rovelli and Karim Thebault. 

\section*{References}

\bdes
\item Achinstein, P. (2001): \textit{The Book of evidence}, Oxford: Oxford University Press.
\item Bovens, L., and Hartmann, S.\ (2003): \textit{ Bayesian Epistemology}, Oxford: Oxford University Press.
\item Dawid, R. (2006): "Underdetermination and Theory Succession from the Perspective of String Theory", Philosophy of Science 73/3, 298-322.
\item Dawid, R. S. Hartmann and Jan Sprenger (2015): "The No Alternatives Argument",  The British Journal for the Philosophy of Science 66(1), 213-234. 
\item Dawid, R. (2013): \textit{String Theory and the Scientific Method}, Cambridge: Cambridge University Press.
\item Ellis, G. and J. Silk (2014): "Defend the Integrity of Physics", Nature 516: 321-323.
\item Howson, C., and Urbach, P.\ [2006]: \textit{Scientific Reasoning: The Bayesian Approach,} third edition, La Salle: Open Court.
\item Rickles, D. (2013): "Mirror Symmetry and other Miracles in Superstring Theory", Foundations of Physics 43(1): 54-80.  
\item Rovelli, C. (2017): "The Dangers of Non-Empirical Confirmation", arXiv:1609.01966.
\item Sklar, L. (1975): "Methodological Conservativism", Philosophical Review 84, 384.
\item Stanford, K. (2006): \textit{Exceeding Our Grasp - Science, History and the Problem of Unconceived Alternatives}, Oxford: Oxford University Press.

\edes

\end{document}